%% file: main.tex
\documentclass[conference]{IEEEtran}
\ifCLASSINFOpdf
\else
\fi
\usepackage{amssymb}
\usepackage{graphicx}
\usepackage{amsmath}
\usepackage{caption}
\usepackage{subcaption}

\usepackage{tikz} 
\usetikzlibrary{arrows,automata,shapes} % LATEX and plain TEX 
\tikzstyle{every picture}+=[remember picture]
\everymath{\displaystyle}

\usepackage{url}
\begin{document}
\title{Sapo: Reachability Computation and Parameter Synthesis of Polynomial Dynamical Systems}

\author{\IEEEauthorblockN{Tommaso Dreossi}
\IEEEauthorblockA{University of California, Berkeley\\
Email: tommasodreossi@berkeley.edu}}


%Telephone: (800) 555--1212, Fax: (888) 555--1212}


\maketitle

\begin{abstract}
Sapo is a C++ tool for the formal analysis of polynomial 
dynamical systems. Its main features are: 1) \emph{Reachability computation},
i.e., the calculation of the set of states reachable from a set of initial conditions,
and 2) \emph{Parameter synthesis}, i.e., the refinement of a set of parameters so that the system satisfies a given specification.
Sapo can represent reachable sets as unions of boxes, parallelotopes,
or parallelotope bundles (symbolic representation of polytopes).
Sets of parameters are represented with polytopes while 
specifications are formalized as Signal Temporal Logic (STL) formulas.
\end{abstract}


% \fi
\IEEEpeerreviewmaketitle

\input{introduction}

\input{reach_parasynth}

\input{main_features}

\input{structure_and_usage}

\input{conclusion}



%\bibliographystyle{IEEEtran}

\input{appendix}

\end{document}

%% file: introduction.tex
%!TEX root = bare_conf.tex

\section{Introduction}\label{sec:introduction}

Formal verification involves the strict and exhaustive
study of systems using mathematically based techniques. The development 
and application of formal methods is motivated by the fact that the formal study of 
a model implies the reliability and robustness of the abstracted system.

Two important problems emerging from the formal analysis of dynamical systems are 
the \emph{reachability computation} and the \emph{parameter synthesis} problems.
In the first case, for a given set of initial conditions,
it is asked to compute the set of states reachable
by the dynamical system. In the second case, an initial set of conditions together 
with a set of parameters and a specification are provided, and it is asked to 
refine the set of parameters in such a way that the system satisfies the specification.
These two problems often recur in 
formal analysis. For instance,
the reachability computation can be used to determine whether
a systems reaches some undesired states or remains within a particular subset of the state space,
while synthesis of parameters can be used to design dynamical systems
and tune their parameters so that to fit
experimental observations or satisfy requirements.

Despite the numerous tools available for the analysis of linear dynamical 
systems (i.e., systems whose dynamics are linear functions), not many tools
for the study of nonlinear systems are available. 
The main difficulties arising from nonlinearity concern the transformation  of sets with respect to
nonlinear functions. This operation, in general, does not preserve nice properties of the 
sets (e.g., convexity), and thus approximation techniques, that often suffer in terms of scalability, are necessary.

In this work we present Sapo~\cite{sapoweb}, a C++ tool that targets discrete-time polynomial dynamical
systems (possibly equipped with
parameters) and that deals with both the
reachability computation and parameter synthesis problems.
Sapo can be used either to generate flowpipes that over-approximate 
reachable sets over bounded time horizons, or to refine sets of parameters in such a way that the system 
satisfies a Signal Temporal Logic (STL)~\cite{maler2004monitoring} specification.
The tool allows the construction of flowpipes using boxes  (i.e., hyperrectangles),
parallelotopes (i.e., $n$-dimensional parallelograms), or parallelotope bundles
(i.e., sets of 
parallelotopes whose intersections symbolically represent polytopes),
and the refinement of parameter sets represented by polytopes.
Sapo's underlying algorithms exploit
some properties of Bernstein coefficients of polynomials. Intuitively, the Bernstein coefficients 
can be used to bound a polynomial 
over the unit box domain (i.e., $[0,1]^n$). The core idea on which Sapo relies is the transformation of 
unit boxes into generic sets. With this trick, Sapo exploits Bernstein coefficients to maximize polynomials and
determine sets that over-approximate the trajectories of the system. A similar approach
is also used to reduce the parameter synthesis problem into linear programs, where the linear systems, 
constructed through parameterized Bernstein coefficients, represent sets of valid parameters.
The algorithms implemented in Sapo are based on the theoretical results exposed
in~\cite{dang2012reachability,DreossiD14,DangDP14,DangDP15,DreossiDP16,dreossi2016thesis}.
%The tool is available at the link \url{http://www.eecs.berkeley.edu/~tommasodreossi/sapo/}.

%% file: reach_parasynth.tex
%!TEX root = bare_conf.tex

\section{Reachability and Parameter Synthesis}\label{sec:reach_parasynth}

Sapo deals with discrete-time parametric polynomial dynamical systems
described by difference equations of the form:
\begin{equation}~\label{eq:dyn}
	x_{k+1} = f(x_k,p)
\end{equation}
where $f : \mathbb{R}^n \times \mathbb{R}^m \to \mathbb{R}^n$ is a
polynomial linear in $p$, $x_k \in \mathbb{R}^n$ is the state of the system at time $k \in \mathbb{N}$,
and $p \in \mathbb{R}^m$ are the parameters. Given an initial condition $x_0 \in \mathbb{R}^n$
and parameters $p \in \mathbb{R}^m$, the trajectory $x_0, x_1, x_2, \dots$ describing the 
evolution the system, can be obtained by iterating the function (\ref{eq:dyn}).
The problems addressed by Sapo are:

\begin{itemize}
	\item \emph{Reachability computation}: Given a set of initial conditions $X_0 \subset \mathbb{R}^n$, 
		a set of parameters $P \subset \mathbb{R}^m$, and a time instant $T \in \mathbb{N}$,
		determine a sequence of tight sets $X_0, X_1, X_2, \dots X_T$, called flowpipe,
		such that all the trajectories of length $T$ with initial conditions in $X_0$ and parameters in $P$ are included in the constructed flowpipe;
	\item \emph{Parameter synthesis}: Given a set of initial conditions $X_0 \subset \mathbb{R}^n$, 
		a set of parameters $P \subset \mathbb{R}^m$, and a specification $\varphi$,
		determine the largest set  $P_\varphi \subseteq P$ such that all the trajectories with
		initial conditions in $X_0$ and parameters in $P_\varphi$ satisfy $\varphi$.
\end{itemize}

%% file: main_features.tex
%!TEX root = bare_conf.tex

\section{Main Features}\label{sec:main_features}

\subsection{Bernstein Coefficients}

At the core of Sapo for both reachability computation and parameter synthesis there are Bernstein coefficients of polynomials.
Intuitively, Bernstein coefficients can be used to represent a polynomial in Bernstein form, i.e., as the linear
combination of Bernstein basis and coefficients. 
Given a polynomial $\pi : \mathbb{R}^n \to \mathbb{R}$:
\begin{equation}
	\pi(x)  = \sum_{i \in I^\pi} a_i x^i
\end{equation}
where $i = (i_1, \dots, i_n) \in \mathbb{N}^n$ is a multi-index,
$x^i = x_1^{i_1}\dots x_n^{i_n}$ is a monomial, $a_i \in \mathbb{R}$ is a coefficient, and $I^\pi \subset \mathbb{N}^n$ is the multi-index set of $\pi$,
the $i$-th \emph{Bernstein coefficient} is:
\begin{equation}
	b_i = \sum_{j \leq i} \frac{\binom{i}{j}}{\binom{d}{j}} a_j
\end{equation}
where $i \leq j$ if $i_k \leq j_k$ for $k=1,\dots,n$, 
$\left( \begin{smallmatrix} i \\ j \end{smallmatrix} \right)$ is the product of the binomial coefficients 
$\left( \begin{smallmatrix} i_1 \\ j_1 \end{smallmatrix} \right) \dots \left( \begin{smallmatrix} i_n \\ j_n \end{smallmatrix} \right)$, and $d$ is the smallest
multi-index such that $i \leq d$ for all $i \in I^\pi$.

One of the interesting properties of Bernstein coefficients is the
\emph{range enclosure property}~\cite{cargo1966bernstein} stating that
$\min_{i \in I^\pi} b_i \leq \pi(x) \leq \max_{i \in I^\pi} b_i$ 
for $x \in [0,1]^n$. This implies that the maximum Bernstein 
coefficient is an upper bound of the maximum of $\pi$
over the unit box domain. Thus, to optimize a polynomial, instead of using
nonlinear/nonconvex optimization techniques, one can compute
the Bernstein coefficients and extract their maximum. The main drawback of this approach 
is that the enclosure property holds only on the unit box domain.

In~\cite{dreossi2016thesis} we extended the range enclosure property to parametric polynomials
of the form $\pi : \mathbb{R}^n \times \mathbb{R}^m \to \mathbb{R}$,
showing that $\min_{i \in I^\pi} \min_{p \in P} b_i(p) \leq \pi(x,p) \leq \max_{i \in I^\pi} \max_{p \in P}b_i(p)$ for all $x \in [0,1]^n$ and parameters $p \in P$.
This property, with a slight adaptation of the treated  polynomial, is the 
key element of Sapo's reachability and parameter synthesis algorithms.

\subsection{Reachability Computation}
The flowpipe $X_0, X_1, \dots, X_T$ that includes the reachable set
can be obtained as a sequence of set image transformations
$X_{k+1} = f(X_k,P)$. Unluckily, this operation becomes difficult when $f$ is nonlinear,
since many properties of sets (e.g., convexity) can be lost.
A common method to circumvent this problem consists in over-approximating the image of sets with
simpler objects such as \emph{polytopes}. A polytope $X \subset \mathbb{R}^n$ is
a convex set that can be represented by the solutions of a system $D x \leq c$,
where $D \in \mathbb{R}^{m \times n}$ and $c \in \mathbb{R}^m$.
The matrix $D$ and vector $c$ are called \emph{template} and \emph{offset},
and the polytope generated by $D$ and $c$ is denoted by $\langle D, c\rangle$.

Given a template $D$, the offset $c$ such that the set $\langle D, c\rangle$ 
over-approximates the set $f(X_k,P)$ can be determined by solving 
optimization problems of the form:
\begin{equation}\label{eq:max_prob}
	c_i \geq \max_{x\in X_k, p\in P} D_i f(x,p)
\end{equation}
for $i=1,\dots,m$. An upper-bound of the maximum of $D_i f(x,p)$ over the 
unit box can be found computing its Bernstein coefficients and 
maximizing them over the set $P$. However, in order to apply 
this technique on generic domains, we need a slight adjustment of the 
treated polynomial.

In~\cite{DreossiD14} we introduced a method that combines Bernstein coefficients
and parameter synthesis on generic box domains. Intuitively, for a given
box $X_k$, we can compute a map $v : [0,1]^n \to X_k$ that transforms
the unit box into $X_k$. It is easy to see that $f(X_k,P) = f(v([0,1]^n),P)$,
which implies that an upper bound of $D_i f(x,p)$, with $x \in X_k$ and $p \in P$,
can be established determining the maximum Bernstein coefficient of $D_i f(v(x),p)$.
Repeating this operation for all the directions of the template $D$, we obtain 
a new offset $c$ that leads to an over-approximating box
$\langle D, c\rangle \supseteq f(X_k,P)$. 

Changing the map $v$ we can alter the shape of the over-approximation set
and define reachability algorithms based on sets 
different from boxes.
For instance, in~\cite{DangDP14} we shown how the map $v$ can be defined to 
transform unit boxes into parallelotopes. In doing so, we defined
a parallelotope-based set image approximation technique that is
more flexible in the choice
of the initial set and preciser than the box-based one.

The transformation of a unit box into a generic polytope is in general difficult. 
However, in~\cite{DreossiDP16} we introduced a new way or representing polytopes as
finite intersections of polytopes. We called these sets \emph{parallelotope bundles}.
Once that a polytope is decomposed in a collection of parallelotopes, we can reason
separately on each of its parallelotopes and, using the Bernstein coefficients, we can 
determine a new parallelotope bundle that over-approximates the image of the 
starting polytope. This method sensibly increases the accuracy of the over-approximation
flowpipes at the cost of a higher number of optimizations.

In summary, for reachability computation, Sapo supports the construction of flowpipes 
based on boxes, parallelotopes, and parallelotope bundles. All these 
objects are represented by Sapo as both solutions of linear systems
(constraint representation) and affine transformations of unit boxes (generator 
representation). Sapo uses either of these representations
depending on the operation to perform on the set.

\subsection{Parameter Synthesis}

Given a set of initial conditions $X_0$, a set of parameters $P$, and a specification $\varphi$,
we want to determine the largest set $P_\varphi \subseteq P$ such that 
the reachable set starting in $f(X_0,P_\varphi)$ satisfies $\varphi$.

Sapo allows the user to formalize the specifications in terms of 
STL formulas~\cite{maler2004monitoring} in positive normal form, i.e., formulas generated
by the following grammar:
\begin{equation}
	\varphi := \sigma \mid \varphi \wedge \varphi \mid \varphi \vee \varphi \mid \varphi \mathcal{U}_I \varphi
\end{equation}
with $\sigma = g(x_1, \dots, x_n) \leq 0$, where $g : \mathbb{R}^n \to \mathbb{R}$,
and $I \subset \mathbb{N}$ is an interval.
%The common temporal operators such always $\mathcal{G}$
%or eventually $\mathcal{F}$ can be defined in the usual way.}

Similarly to the reachability computation, also the parameter synthesis
algorithm implemented in Sapo exploits the Bernstein coefficients.
Given the sets $X_k,P$ and a predicate $\sigma = g(x_1, \dots, x_n) \leq 0$,
the set $P$ is a valid parameter set if $g(f(x,p)) \leq 0$ holds for all $x \in X_k$ and $p \in P$.
This constraint can be verified by determining a map $v : [0,1]^n \to X_k$,
computing the Bernstein coefficients of the function $g(f(v(x),p))$, and 
checking that $b_i(p) \leq 0$ for all $b_i(p) \in I^{g \circ f \circ v}$ and $p \in P$.
In particular, a refinement $P_\sigma$ of the set $P$ with respect to 
the predicate $\sigma$ can be obtained by adding the linear constraints 
$b_i(p) \leq 0$ to the linear system representing $P$. The solutions 
of the new linear system are the valid parameters. This
condition can be verified solving a single linear program.

Reasoning by structural induction on the given specification, it is 
possible to reduce the parameter synthesis problem into a 
collection of refinements over predicates. The partial results are
then combined accordingly with the treated formula. The refinement
of the until formula requires also the computation of the evolution of the system,
task that can be achieved by the reachability methods previously introduced.
The parameter synthesis algorithm, its correctness, and its complexity are 
presented in detail in~\cite{DangDP15,dreossi2016thesis}.

%The sets of parameters supported by Sapo are polytopes represented as 
%solutions of linear systems. For a given set of initial conditions, parameter set 
%and time instant, Sapo produces a sequence of sets of the same kind of the initial 
%one that constitutes an over-approximation of the reachable set of the dynamical 
%systems. The tool also handles nonparametric dynamical systems for which it 
%is not required to specify an initial parameter set.

In summary, for the parameter synthesis, Sapo receives in input a set of initial 
conditions representable with a box or a parallelotope, a polytopic set of parameters specified
as a linear system, an STL formula, and it produces a collection of 
polytopes representing the set of parameter set under which the system satisfies the specification.

%% file: structure_and_usage.tex
%!TEX root = bare_conf.tex

\section{Structure and Usage}\label{sec:structure_and_usage}

\subsection{Tool Architecture}

% Define block styles
\tikzstyle{decision} = [diamond, draw, fill=blue!20, 
    text width=4.5em, text badly centered, node distance=3cm, inner sep=0pt]
\tikzstyle{block} = [rectangle, draw, 
    text width=4.8em, text centered, rounded corners, minimum height=4em]
\tikzstyle{line} = [draw, -latex']
\tikzstyle{cloud} = [draw, ellipse,fill=red!20, node distance=3cm,
    minimum height=2em]
   
\begin{figure*}
\centering
\resizebox{.65\textwidth}{!}{
\begin{tikzpicture}[node distance = 2.25cm, auto]
    % Place nodes
    \node [block] (sapo) {\small Sapo core};
    \node [block, below of=sapo] (base_converter) {\small Base converter};
    \node [block, right of=base_converter, node distance=2.75cm] (bundle) {\small Bundle};
    \node [block, right of=bundle, node distance=2.75cm] (lsset) {\small Linear system set};
    \node [block, left of=base_converter, node distance=2.75cm] (model) {\small Model};
    \node [block, left of=model, node distance=2.75cm] (stl) {\small STL formula};
    \node [block, below of=bundle] (paratope) {\footnotesize Parallelotope};
    \node [block, below of=lsset] (ls) {\small Linear system};	
    %\node [block, below of=paratope] (varsgen) {\small Vars Generator};

    % Draw edges
    \path [line] (base_converter) -- (sapo);
    \path [line] (stl) |- (sapo);
    \path [line] (model) -- (sapo);
    \path [line] (bundle) -- (sapo);
    \path [line] (lsset) |- (sapo);
    \path [line] (paratope) -- (bundle);
    %\path [line] (varsgen) -- (paratope);
    \path [line] (ls) -- (lsset);
    \path [line] (ls) -- (paratope);
    \path [line] (ls) -- (bundle);
    \path [line] (bundle) -- (base_converter);
    
    \draw[black, dashed, rounded corners=.15cm] (-6.75,-5.5) rectangle (6.75,1);
     \node at (6,0.5) {Sapo};
    
\end{tikzpicture} 
}
\caption{Architecture of Sapo.}
\label{fig:sapo_struct}
\end{figure*}
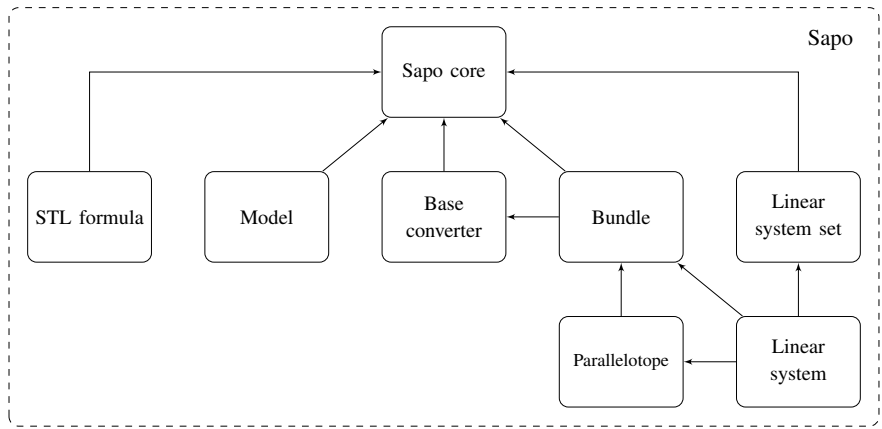

Sapo is implemented in C++. It relies on the libraries GiNaC\footnote{\url{http://www.ginac.de}}
for symbolically manipulating polynomials
and on GLPK\footnote{\url{https://www.gnu.org/software/glpk/}}
for solving linear programs.

Figure~\ref{fig:sapo_struct} summarizes the structure of the tool. The user can specify a
dynamical system and a possible specification using the \texttt{Model} and \texttt{STL}
modules, respectively. The set of initial conditions and possible parameters can be specified
using the modules \texttt{Bundle} and \texttt{LinearSystemSet}, respectively (a single box or parallelotope can be seen as a bundle with a single template). With these elements, the user 
can analyze the dynamical system invoking the main module \texttt{SapoCore}. This block
implements the reachability and parameter synthesis algorithms and returns
the computed results. An important module with which \texttt{SapoCore} interacts is the 
\texttt{BaseConverter}. This block computes
Bernstein coefficients of polynomial functions heavily exploited by our algorithms.
The \texttt{BaseConverter} symbolically computes the 
Bernstein coefficients of a polynomial only once.
The symbolic coefficients are stored in a data structure from which they are fetched and numerically instantiated whenever needed. This trick allows the tool to save computations
and sensibly speeds up the analysis. Bernstein coefficients are
computed by 
default using our improved matrix method~\cite{DreossiD14}.
%Finally, the module \texttt{VarsGenerator}
%is used to generate the variables appearing in the generator representation of sets.
To visualize the results, Sapo gives the possibility to generate a Matlab script
that displays 2/3-dimensional sets or projections of higher dimensional sets.

\subsection{Experimental Evaluation}

As a demonstration, we apply Sapo to the SIR epidemic model, a well-known 3d nonlinear system that describes the progress of a disease in a 
population.
%The dynamics of the SIR model are:
%\begin{equation}
%	\begin{split}
%		s_{k+1} = &\ s_k - (\beta s_k i_k) \Delta \\
%		i_{k+1} = &\ i_k + (\beta s_k i_k - \gamma i_k) \Delta \\
%		r_{k+1} = &\ r_k + (\gamma i_k) \Delta \\
%	\end{split}
%\end{equation}
The model considers three groups of individuals: $s$ the susceptible healthy,
$i$ the infected, and $r$ the removed from the system (e.g., the recovered ones).
Two parameters regulate the system's evolution: $\beta$ the contraction rate and $\gamma$,
where $1/\gamma$ is the mean infective period.

\begin{figure}
	\centering
	\begin{subfigure}{0.2\textwidth}
		\includegraphics[scale=0.08]{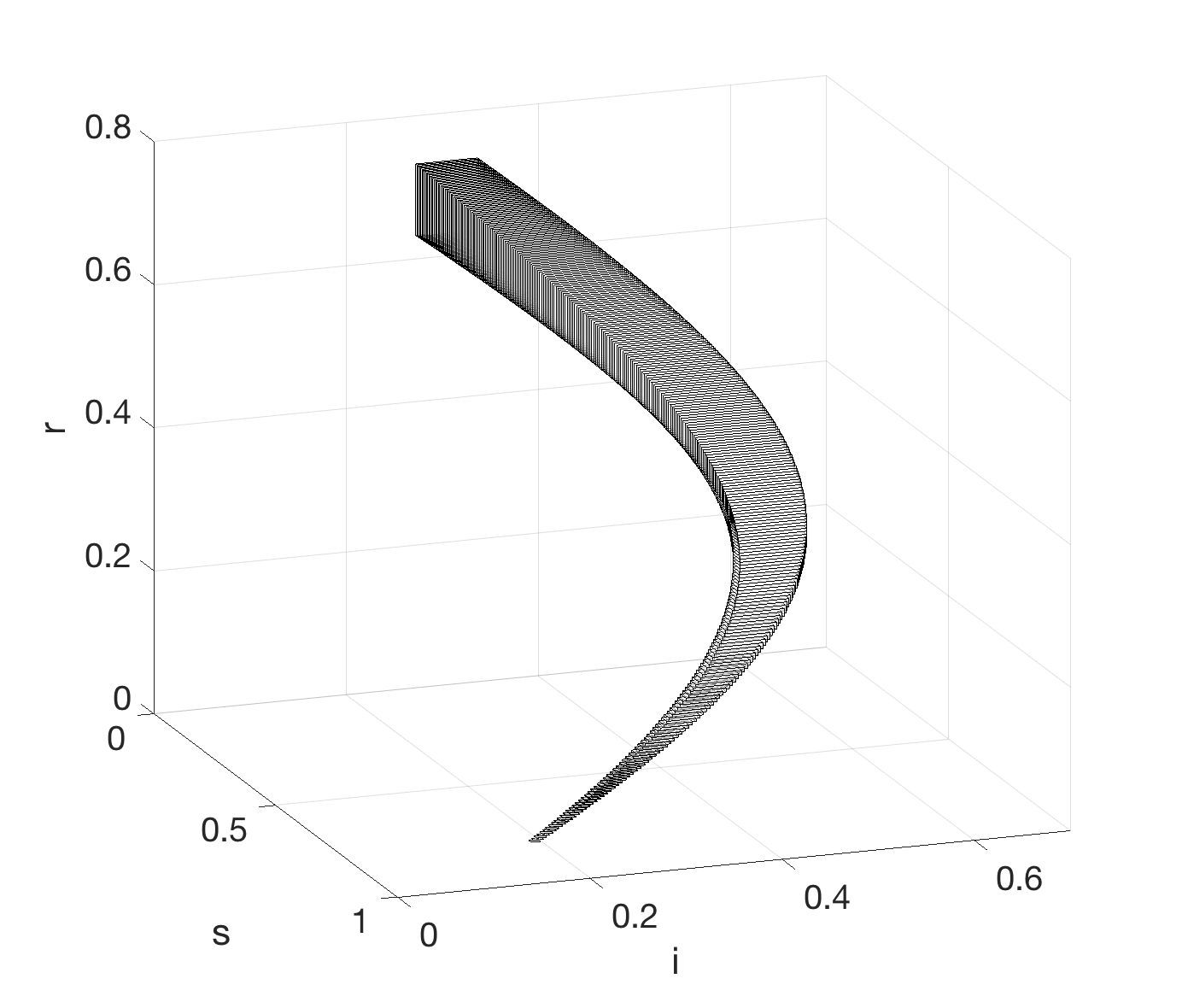}
		\caption{Box-based\label{fig:box_reach_set}}
	\end{subfigure}
	\begin{subfigure}{0.2\textwidth}
		\includegraphics[scale=0.082]{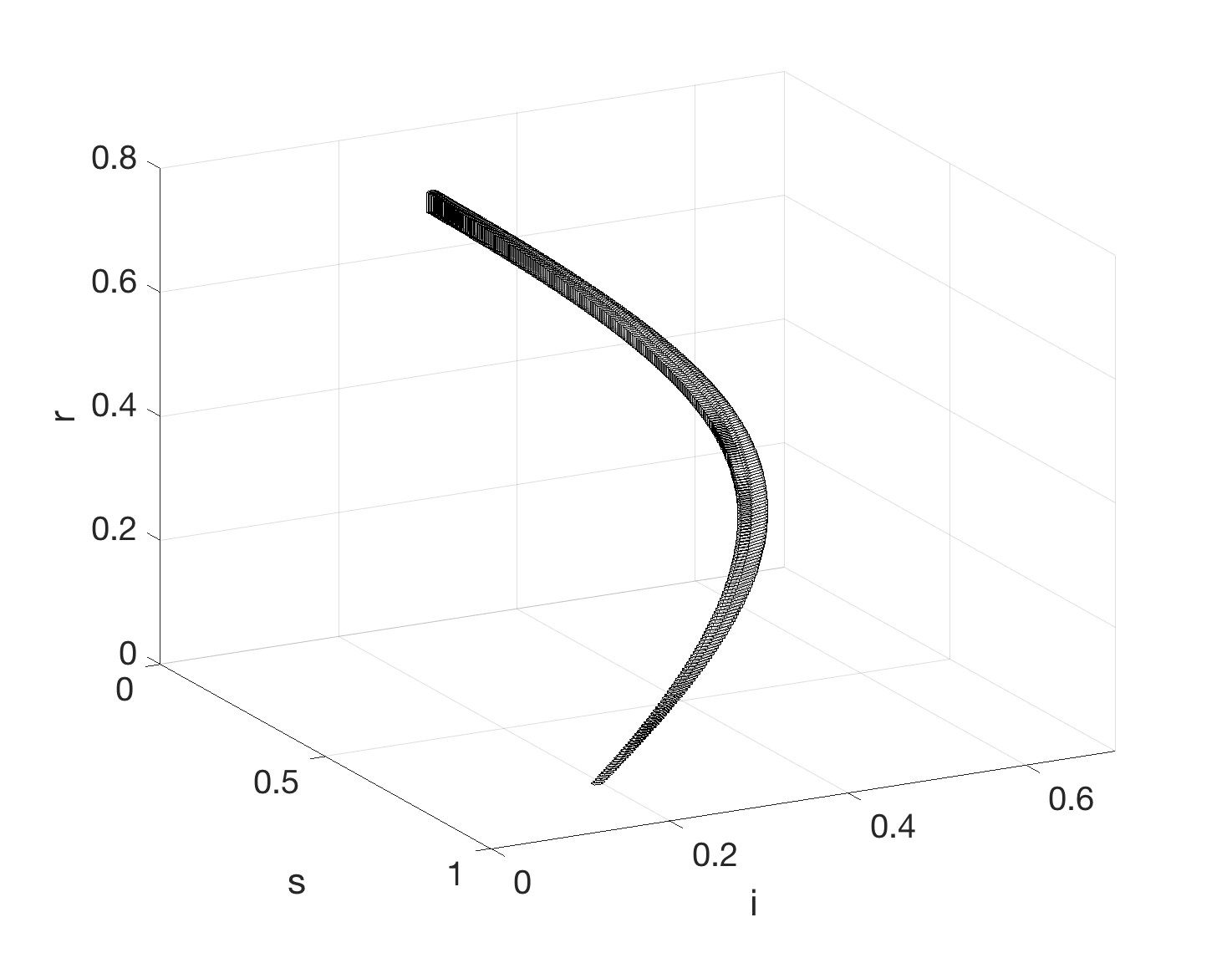}
		\caption{Bundle-based\label{fig:bund_reach_set}}
	\end{subfigure}
	\caption{Reachable sets of SIR model (300 steps)\label{fig:reach_set}}
\end{figure}

We used Sapo to compute the bounded time reachable set of the SIR model with different configurations.
At first, with parameters $\beta = 0.34, \gamma = 0.05$, and set of initial conditions
$s_0 \in [0.79, 0.80], i_0 \in [0.19, 0.20]$, and $r_0 \in [0.00, 0.00]$,
we computed in $0.12$s the reachable set for 300 steps using a single box (see Figure~\ref{fig:box_reach_set}).
Then, by adding two surrounding
parallelotopes to the set of initial conditions, we obtained in $2.83$s a bundle-based flowpipe that better over-approximates the 
reachable set (see Figure~\ref{fig:bund_reach_set}).
Notice how, for the same initial set, the bundle-based
flowpipe is more accurate than the box-based one.

\begin{figure}
	\centering
	\begin{subfigure}{0.2\textwidth}
		\includegraphics[scale=0.085]{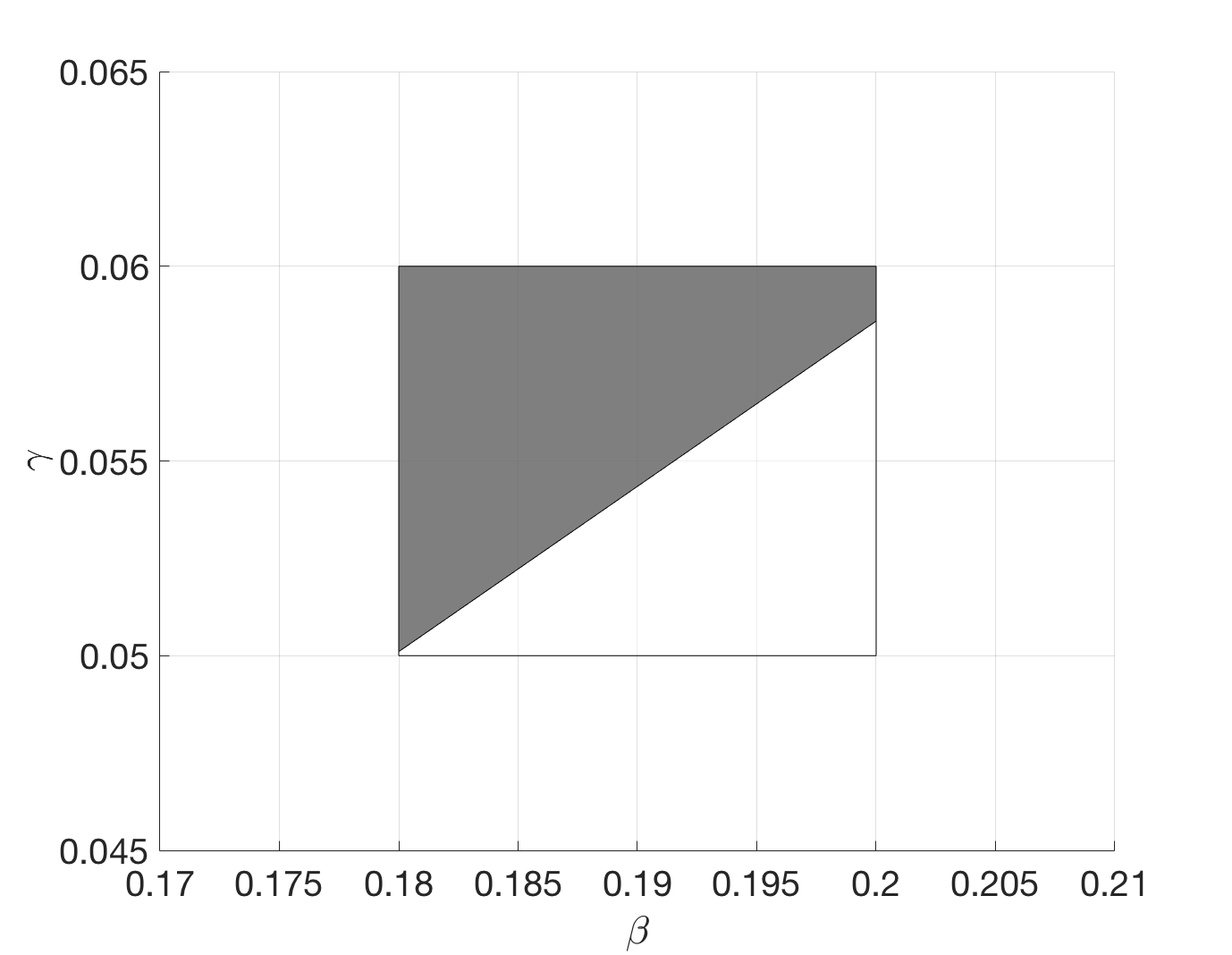}
		\caption{Synthesized parameters\label{fig:para_set}}
	\end{subfigure}
	\begin{subfigure}{0.2\textwidth}
		\includegraphics[scale=0.075]{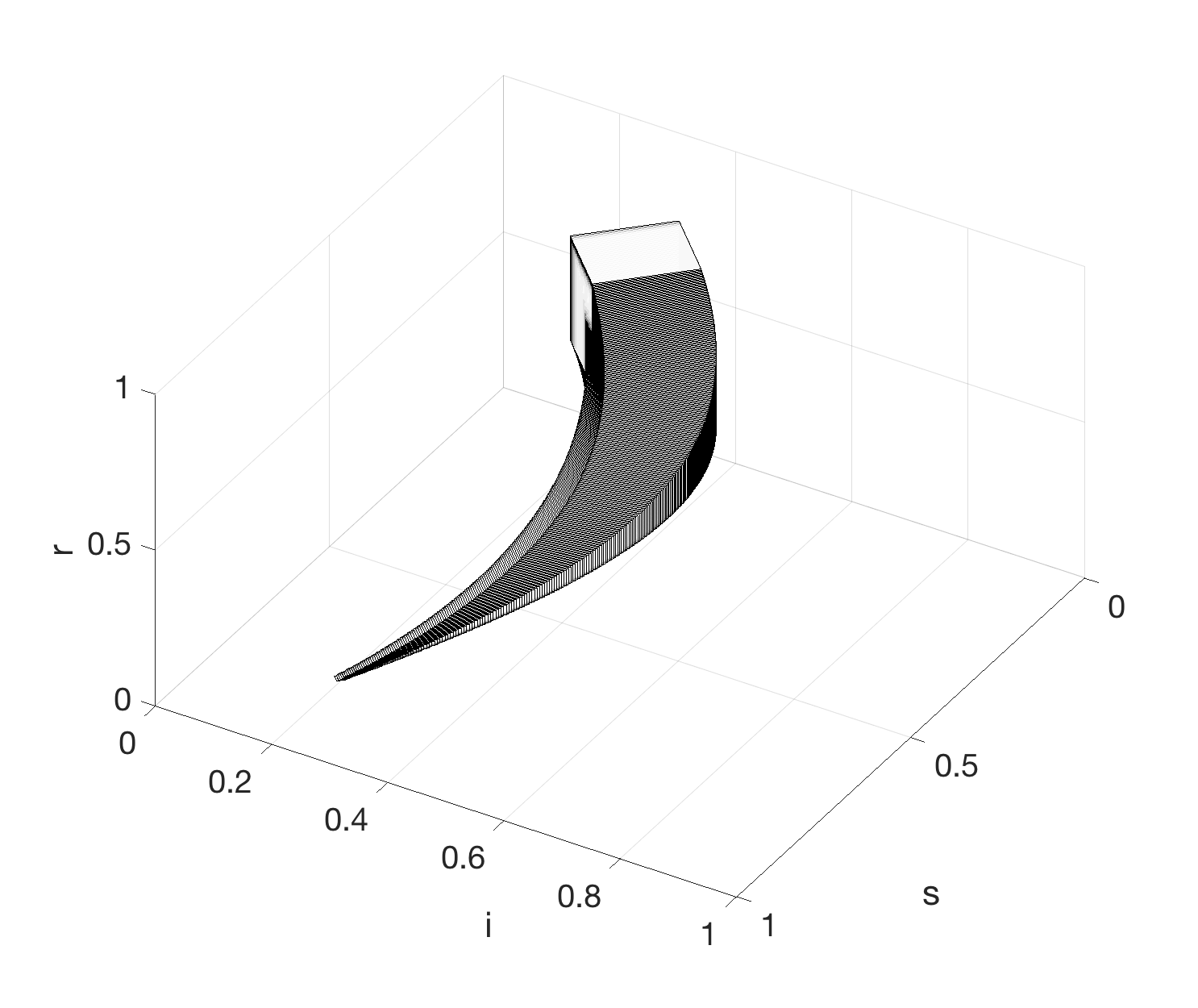}
		\caption{Constrained evolution\label{fig:para_reach_set}}
	\end{subfigure}
	\caption{Parameter synthesis of SIR model ($G_{[50,100]}(i \leq 0.44)$)\label{fig:para_synth}}
\end{figure}

Sapo can also be used to synthesize sets of valid parameters.
For instance, considering the set of initial conditions $s_0 \in [0.79,0.80]$, $i_0 \in [0.19,0.20]$, and
$r_0 = [0.00,0.00]$, and the initial set of parameters $\beta \in [0.18,0.20]$ and $\gamma \in [0.05,0.06]$,
we can ask Sapo to refine the parameter set so that the STL formula $G_{[50,100]}(i \leq 0.44)$ holds. 
Figure~\ref{fig:para_set} shows the original parameter set (in white)
and the synthesized one (in gray) computed in $0.83$s, while Figure~\ref{fig:para_reach_set}
depicts the evolution of the system under the synthesized 
parameter set.
%For details on the scalability of our algorithms the reader may refer to~\cite{DangDP15,dreossi2016thesis}.

We evaluated both Sapo's reachability and parameter synthesis algorithms on several dynamical systems~\cite{DreossiD14, DangDP14, DangDP15, dreossi2016thesis, DreossiDP16}.
We computed the reachable sets of biological models, such as SIR (3d) or a honeybees nest choice model (5d),
or of a quadrotor drone (17d) (see Appendix~\ref{sec:reach_bench}), and synthesized
parameters of epidemic systems, such as SIR (3d, 2params), influenza (4d, 2params),
Ebola (5d, 4params), and SARS (6d, 4params) (see Appendix~\ref{sec:synth_bench}).

%% file: conclusion.tex
%!TEX root = bare_conf.tex

\section{Conclusion}\label{sec:conclusion}

\subsection{Related Work}

	In the last two decades numerous tools have been developed for the 
	reachability analysis of linear dynamical 
	system and hybrid automata. Some examples 
	are SpaceEx~\cite{frehse2011spaceex}, HyTech~\cite{henzinger1997hytech}, or d/dt~\cite{asarin2002d}.
	Unlike for the linear case, not many tools for the nonlinear 
	analysis are available. Some tools, such as 
	Breach~\cite{donze2010breach}, S-TaLiRo~\cite{annpureddy2011s}, or C2E2~\cite{duggirala2015c2e2}
	address nonlinearity using a finite number of simulations.
	Tools that directly deal with nonlinear reachability
	are dReach~\cite{kong2015dreach} and Flow*\cite{chen2013flow}. The latter is probably the
	closest tool to Sapo in terms of both problem formulation and 
	provided results. The main difference between Flow* and Sapo is that
	the first produces flowpipes grouping collections of Taylor models,
	while the second produces a collection of polytopes computed using Bernestein 
	coefficients. Another difference is that Flow* considers hybrid automata,
	while Sapo, at the moment, focuses only on dynamical systems.	 

\subsection{Future Developments}

	Sapo is still under development. As future work, we will extend the reachability
	computation to polynomial hybrid automata, where invariants and guards need to be 
	considered. An interesting aspect of operations on parallelotope bundles is that 
	they can be easily parallelized. It could be interesting to implement a parallel version 
	of the tool and investigate its scalability. Finally, a comparison with the 
	available tools for nonlinear reachability is on our schedule.

%% file: appendix.tex
%!TEX root = bare_conf.tex

\begin{appendices}

	\section{Reachability}\label{sec:reach_bench}
	
		\begin{table}	
		\begin{center}
			\begin{tabular}{| c | c c c | c |}
				\hline
				Model	&	Vars	&	Dir/Temps		&	 Steps	&	Time	\\ 
				\hline
				SIR~\cite{kermack1927contribution}		&	2	&	4/2			&	25		&	0.24		\\
						&		&	3/1			&	60		&	0.12 		\\
						&		&	6/4			&	60		&	2.83		\\
				Nest choice~\cite{britton2002deciding}	&	5	&	5/1			&	1500		&	26.90	\\ 
						&		&	7/3			&	1500		&	81.27	\\
			Quadcopter~\cite{edo2015arch}	&	17	&	17/1			&	300		&	17.74	\\
						&		&	18/2			&	300		&	39.07	\\
				\hline
			\end{tabular}
		\end{center}
		\caption{Reachability computation benchmark\label{tab:reach_bench}}
		\end{table}
		
		Table~\ref{tab:reach_bench} reports the evaluation of Sapo's reachability computation algorithm.
		The tool has been evaluated on three nonlinear systems: the SIR epidemic model~\cite{kermack1927contribution},
		a honeybees site choice model~\cite{britton2002deciding} 
		describing the mechanism adopted by a swarm of honeybees to choose
		one among two different nest-sites, and a quadrotor drone model~\cite{edo2015arch} characterizing both the drone and its controller dynamics. 
		Each system has been tested on different combinations of number of directions and templates.
		The table reports the dimension of the system (i.e., the number of variables), the number of directions and templates used for the
		parallelotope bundle, the total number of reachability steps, and the computation times expressed in seconds.

	\section{Parameter Synthesis}\label{sec:synth_bench}
	
		The scalability of the parameter synthesis algorithm has been studied in 
		terms of system dimension (Section~\ref{sec:synth_1})
		and specification's length (Section~\ref{sec:synth_2}).
	
		\subsection{Benchmark 1}\label{sec:synth_1}
			
		\begin{table}	
		\begin{center}
			\begin{tabular}{| c | c c c | c |}
				\hline
				Model	&	Vars		&	 Params	&	 Spec &	 Time	\\
				\hline
				SIR~\cite{kermack1927contribution}		&	3		&	2		&	$\mathcal{G}_{[10,30]}(i \leq 0.68)$	&	0.10 \\
						&			&			&	$\mathcal{G}_{[20,50]}(i \leq 0.50)$	&	0.65 \\
				Influenza~\cite{gonzalez2011note}	&	4		&	2		&	$\mathcal{G}_{[0,50]}(i \leq 0.43)$	&	6.27 \\
				Ebola~\cite{chowell2004basic}	&	5		&	4		&	$(i \leq 200)\mathcal{U}_{[6,10}(q \leq 20)$	&	0.10 \\
										&			&			&  	$(q \leq 40)\mathcal{U}_{[10,15]}(i \leq 270)$	& 	0.14 \\
										&			&			&  	$(q \leq 50)\mathcal{U}_{[5,15]}(e > 100 \vee q > 25)$	& 	0.14 \\										
											
				\hline
			\end{tabular}
		\end{center}
		\caption{Parameter synthesis benchmark (scalability in system's dimension) \label{tab:synth_dim}}
		\end{table}
		
		Table~\ref{tab:synth_dim} reports the evaluation
		considering models of increasing dimension.
		We considered three nonlinear dynamical systems: the SIR model~\cite{kermack1927contribution},
		a simplification of the influenza model presented in~\cite{gonzalez2011note},
		and a variation of the model describing the Ebola outbreak in Congo 1995 and Uganda 2000
		exposed in~\cite{chowell2004basic}.
		For each model we considered different specifications.
		The table reports the dimension of the system, the numbers of synthesized parameters, the considered STL specification, 
		and the computation time expressed in seconds.	

\subsection{Benchmark 2}\label{sec:synth_2}

\begin{table}
  \begin{center}
  \begin{subtable}[b]{0.4\textwidth}
  \begin{center}
   \begin{tabular}{| c c | c c c |}
   \multicolumn{4}{ c }{}\\[0.375cm]
   \hline
    $a$ & $b$ & $\phi_1$ & $\phi_2$ & $\phi_3$\\
   \hline
    5 & 15 & 0.20 (11) & 0.15 (7) & 0.14 (4)\\
    5 & 20 & 0.35 (16) & 0.24 (11) & 0.21 (4)\\
    5 & 30 & 0.82 (26) & 0.55 (21) & 0.36 (4)\\
    5 & 50 & 2.63 (46) & 1.65 (41) & 0.80 (4)\\
    5 & 100 & - & 10.95 (91) & 1.69 (4)\\
    5 & 125 & - & - & 1.69 (4)\\
    15 & 20 & 0.29 (6) & 0.22 (4) & 0.19 (0)\\
    20 & 30 & 0.64 (11) & 0.45 (11)& 0.35 (0)\\
    30 & 50 & 1.88 (21)& 1.27 (21) & 0.78 (0)\\
    50 & 100 & 13.00 (51) & 6.89 (51) & 1.72 (0)\\
    100 & 200 & - & -   & 1.66 (0)\\
    \hline
    \multicolumn{4}{ c }{}\\
    \multicolumn{4}{ c }{\hspace{1cm}(a) Increasing until interval.}    
  \end{tabular}
  \end{center}
  \end{subtable}
\\[0.5cm]
  \begin{subtable}[b]{0.4\textwidth}
  \begin{center}
  \begin{tabular}{| c | c c c |}
   \hline
    N & $\phi^N_1$ & $\phi^N_2$ & $\phi^N_3$\\
   \hline
    1 & 0.11 (5) & 0.14 (2) & 0.13 (4) \\
    2 & 0.26 (9) & 0.36 (7) & 0.29 (3) \\
    3 & 0.48 (13) & 0.69 (12) & 0.50 (3) \\
    4 & 0.74 (17) & 1.09 (17) & 0.70 (3) \\
    5 & 1.10 (21) & 1.61 (22) & 1.01 (3) \\
    6 & 1.50 (25) & 2.28 (27) & 1.20 (3) \\
    7 & 1.97 (29) & 3.05 (32) & 1.65 (3) \\
    8 & 2.59 (33) & 3.97 (37) & 1.81 (3) \\
    9 & 3.23 (37) & 5.06 (42) & 2.16 (3) \\
    10 & 4.98 (42) & 6.70 (47) & 2.56 (3) \\
    15 & 10.33 (61) & 11.80 (62) &  4.85 (3) \\
    16 & 11.75 (65) & 15.98 (67) &  5.43 (3) \\
    17 & - & - & 5.97 (3) \\
   \hline
   \multicolumn{4}{ c }{}\\
    \multicolumn{4}{ c }{\hspace{0cm}(b) Nesting until.}    
  \end{tabular}
  \end{center}
  \end{subtable}
  \end{center}
  \caption{Parameter synthesis benchmark (scalability in specification's size). Times and computed polytopes per refinement.
  $\phi_1  \equiv (i \leq 200)\mathcal{U}_{[a,b]} (q \leq 20)$, $\phi_2 \equiv (q \leq 40)\mathcal{U}_{[a,b]} (i \leq 270)$,
  $\phi_3 \equiv (q \leq 50)\mathcal{U}_{[a,b]} (e > 100 \vee q > 25)$.}
  \label{tab:scalability}
\end{table}

For the second evaluation of the parameter synthesis algorithm we considered
the six dimensional Ebola model~\cite{chowell2004basic} and we synthesize
four of its controllable parameters over different specifications.
Let $\phi_1  \equiv (i \leq 200)\mathcal{U}_{[a,b]} (q \leq 20)$, $\phi_2 \equiv (q \leq 40)\mathcal{U}_{[a,b]} (i \leq 270)$,
and $\phi_3 \equiv (q \leq 50)\mathcal{U}_{[a,b]} (e > 100 \vee q > 25)$ be three STL specifications.
Table~\ref{tab:scalability}a reports the evaluations when stretching the time intervals $[a,b]$ of the specifications $\phi_1, \phi_2$, and $\phi_3$.
As second evaluation, we nested several until on the (most critical) right hand side. For instance, the double 
nesting of $\phi_1$ is 
$\phi^N_1  \equiv (I(t) \leq 200)\mathcal{U}{[6,10]} ((I(t) \leq 200)\mathcal{U}{[6,10]} (Q(t) \leq 20))$ with $N=2$.
Table~\ref{tab:scalability}b reports the running times and the number of obtained parameter sets.

\end{appendices}